\documentclass[12pt]{article}

\usepackage{fancybox}
\usepackage{cite}
\usepackage{float}
\usepackage{amsfonts}
\usepackage{amsmath}
\usepackage{amsbsy}
\usepackage{graphicx}
\usepackage{amssymb}
\usepackage{amsthm}
\usepackage{bm}
\usepackage{epsfig}
\usepackage{latexsym}
\usepackage{pdflscape}
\usepackage{color}
\usepackage{here}
\usepackage{graphicx}
\numberwithin{equation}{section}

\allowdisplaybreaks

\DeclareMathOperator{\llangle}{\langle\!\langle\!}
\DeclareMathOperator{\rrangle}{\!\rangle\!\rangle}
\DeclareMathOperator{\im}{{\rm Im}}

\begin{document}

\renewcommand{\thefootnote}{\fnsymbol{footnote}}

\begin{titlepage}
\begin{flushright}
{\footnotesize OCU-PHYS 452}
\end{flushright}
\bigskip
\begin{center}
{\LARGE\bf Dualities in ABJM Matrix Model\\[8pt]
from Closed String Viewpoint
}\\
\bigskip\bigskip
{\large 
Kazuki Kiyoshige\footnote{\tt kiyoshig@sci.osaka-cu.ac.jp}
\quad and \quad
Sanefumi Moriyama\footnote{\tt moriyama@sci.osaka-cu.ac.jp}
}\\
\bigskip
{\it Department of Physics, Graduate School of Science,
Osaka City University\\
3-3-138 Sugimoto, Sumiyoshi, Osaka 558-8585, Japan}
\end{center}

\bigskip

\begin{abstract}
We propose a new formalism to study the ABJM matrix model.
Contrary to expressing the fractional brane background with the Wilson loops in the open string formalism, we formulate the Wilson loop expectation value from the viewpoint of the closed string background.
With this new formalism, we can prove some duality relations in the matrix model.
\end{abstract}

\bigskip\bigskip
\centering\includegraphics[scale=0.8]{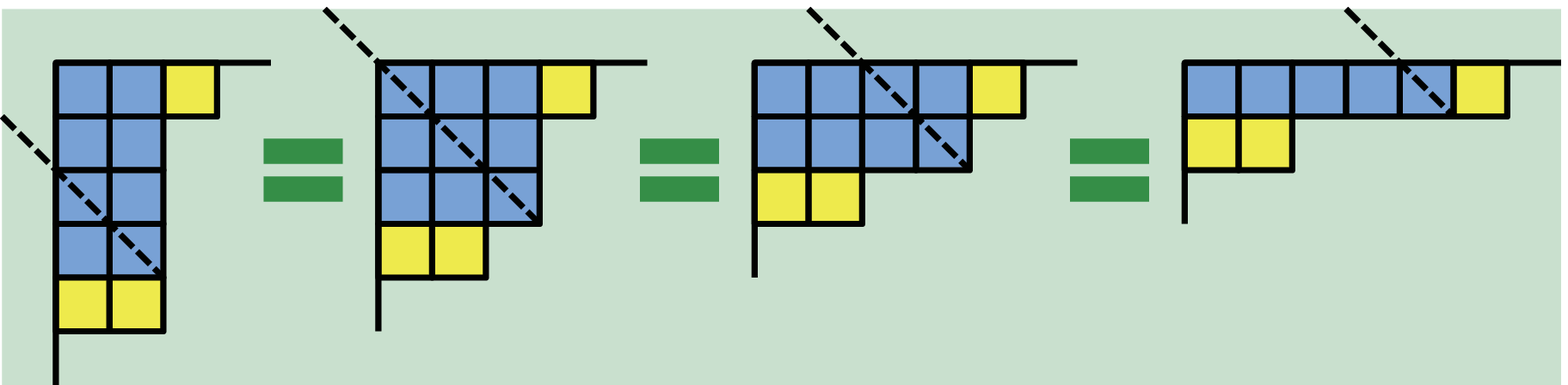}
\end{titlepage}


\renewcommand{\thefootnote}{\arabic{footnote}}
\setcounter{footnote}{0}

\section{Introduction and summary}\label{intro}

The most fundamental theory in theoretical physics is probably M-theory, which is an eleven-dimensional theory considered to unify all of the five ten-dimensional perturbative string theories.
M2-branes and M5-branes are respectively fundamental and solitonic excitations in M-theory.
From the fundamental roles it plays in theoretical physics, we naturally expect a large number of duality relations in M-theory.
However, it is difficult to observe these dualities directly, since only little is known for this mysterious theory.

Recently, it was proposed \cite{ABJM,HLLLP2,ABJ} that the ${\mathcal N}=6$ superconformal Chern-Simons theory with gauge group U$(N_1)\times$U$(N_2)$, Chern-Simons levels $(k,-k)$ and two pairs of bifundamental matters describes the worldvolume theory of coincident $\min(N_1,N_2)$ M2-branes and $|N_2-N_1|$ fractional M2-branes on a geometry ${\mathbb C}^4/{\mathbb Z}_k$.
The partition function and vacuum expectation values of the half-BPS Wilson loop on $S^3$, originally defined with the infinite-dimensional path integral, is reduced to a finite-dimensional matrix model \cite{KWY}
\begin{align}
\langle s_Y\rangle_{k}(N_1,N_2)
&=\frac{(-1)^{\frac{1}{2}N_1(N_1-1)+\frac{1}{2}N_2(N_2-1)}}{N_1!N_2!}
\int_{{\mathbb R}^{N_1+N_2}}
\frac{d^{N_1}\mu}{(2\pi)^{N_1}}\frac{d^{N_2}\nu}{(2\pi)^{N_2}}
e^{\frac{ik}{4\pi}(\sum_{m=1}^{N_1}\mu_m^2-\sum_{n=1}^{N_2}\nu_n^2)}\nonumber\\
&\qquad\times\biggl[\frac{\prod_{m<m'}^{N_1}2\sinh\frac{\mu_m-\mu_{m'}}{2}
\prod_{n<n'}^{N_2}2\sinh\frac{\nu_n-\nu_{n'}}{2}}
{\prod_{m=1}^{N_1}\prod_{n=1}^{N_2}2\cosh\frac{\mu_m-\nu_n}{2}}\biggr]^2
s_Y(e^\mu|e^\nu),
\label{vev}
\end{align}
and a hidden super gauge group U$(N_1|N_2)$ was observed \cite{GW,HLLLP2,ABJ,KWY,DT}.
Here $s_Y(x|y)$ is the supersymmetric Schur polynomial or the character of U$(N_1|N_2)$.
We directly observe a relation under the exchange of two sets of integration variables,
\begin{align}
\langle s_Y\rangle_{k}(N_1,N_2)=\langle s_{Y^\text{T}}\rangle_{-k}(N_2,N_1)
=[\langle s_{Y^\text{T}}\rangle_{k}(N_2,N_1)]^*,
\label{conj}
\end{align}
because of $s_Y(x|y)=s_{Y^\text{T}}(y|x)$.
Hereafter we shall fix $k>0$.
Also, we often consider the case of $M=N_2-N_1\ge 0$ unless otherwise stated and denote the expectation value $\langle s_Y\rangle_{k}(N,M+N)$ as $\langle s_Y\rangle_{k,M}(N)$.

It was then exciting to find that the partition function (\cite{MPtop,MP,HMO2,HMO3,CM,HMMO} for the case of equal ranks and \cite{sho,HoOk} for non-equal ranks) and vacuum expectation values of the half-BPS Wilson loop \cite{KMSS,HHMO} are respectively expressed in terms of the free energy of the closed and open topological string theories on local ${\mathbb P}^1\times{\mathbb P}^1$, which implies a certain modular invariance. 
Aside from the original computations in the 't Hooft expansion \cite{DMP1,DMP2,FHM}, an important approach that leads to these findings is to rewrite the matrix model into the partition function of a Fermi gas system with $N$ non-interacting particles whose dynamics is governed by a non-trivial density matrix \cite{MP}.
The success in formulating the partition function in terms of that of the Fermi gas system leads to a vast amount of WKB small $k$ expansions \cite{MP,CM} and numerical computations for finite $k$ \cite{KEK,HMO1,PY,HMO2}, from which the relation to the topological strings was found \cite{MPtop,HMO2,HMMO}.

For the partition function in a general background with $M=N_2-N_1$ fractional branes, it was found that
\begin{align}
\frac{\langle 1\rangle_{k,M}(N)}{\langle 1\rangle_{k,M}(0)}
=\frac{1}{N!}\int\frac{d^Nx}{(4\pi k)^N}
\prod_{i<j}^N\biggl(\tanh\frac{x_i-x_j}{2k}\biggr)^2\prod_{i=1}^NV_M(x_i),
\label{pf}
\end{align}
where $V_M(x)$ is defined as
\begin{align}
V_M(x)=\frac{1}{2\cosh\frac{x}{2}}
\prod_{\overline l\in\overline L}\tanh\frac{x+2\pi i\overline l}{2k},
\label{pfV}
\end{align}
with $\overline L=\{M-\frac{1}{2},M-\frac{3}{2},\cdots,\frac{3}{2},\frac{1}{2}\}$.
If we introduce the coordinate and momentum operators satisfying the canonical commutation relation $[\widehat q,\widehat p]=i\hbar$ with $\hbar=2\pi k$, the grand canonical partition function $\langle 1\rangle^\text{GC}_{k,M}(z)=\sum_{N=0}^\infty z^{N}\langle 1\rangle_{k,M}(N)$ is expressed as
\begin{align}
\frac{\langle 1\rangle^\text{GC}_{k,M}(z)}{\langle 1\rangle_{k,M}(0)}
=\det(1+z\widehat\rho_M),
\label{detrho}
\end{align}
with the density matrix $\widehat\rho_M$ given by
\begin{align}
\widehat\rho_M=\sqrt{V_M(\widehat q)}
\frac{1}{2\cosh\frac{\widehat p}{2}}\sqrt{V_M(\widehat q)}.
\label{rho}
\end{align}
The expression \eqref{pf} was first found for the case of equal ranks $M=0$ in \cite{MP} and later extended to the case of non-equal ranks.
Namely, for $M\ne 0$, \eqref{pf} was originally conjectured in \cite{AHS} and proved in \cite{Ho1} with several steps of integrations.
In this paper we shall rederive the result with a more refined presentation motivated by \cite{MS2,MN5} as a byproduct of our analysis.

In the expansion of the determinant in \eqref{detrho} there appear many traces of powers of the density matrix $\widehat\rho_M$.
In introducing a background with $M$ fractional branes, all we have to do is to modify the density matrix $\widehat\rho_M$ \eqref{rho} by changing $V_{M=0}(x)$ into $V_{M}(x)$ \eqref{pfV} without touching the structure of the determinant.
In other words, we express the fractional branes by dressing the density matrix so that the background where the closed strings propagate is changed.
From this viewpoint, this formalism is called ``closed string formalism'' in \cite{PTEP}.

Since some poles of $(2\cosh\frac{x}{2})^{-1}$ in \eqref{pfV} at $x\in 2\pi i({\mathbb Z}\pm\frac{1}{2})$ are cancelled by the zeros of the hyperbolic tangent functions at $x=-2\pi i\overline l$, we can shift the integration contour by $-M\pi i$  \cite{HoOk}.
This shift is essential to make contact with the orthosymplectic Chern-Simons matrix model \cite{MePu,MS1,Ho2,Ok2,MS2,MN5}.
This matrix model is obtained from the localization of the ${\mathcal N}=5$ superconformal Chern-Simons theory with the orthosymplectic gauge group \cite{HLLLP2,ABJ} and the physical interpretation is the introduction of the orientifold plane in the type IIB setup.
In \cite{Ho2,MS2,MN5} it was proved\footnote{The proof for odd $M$ is motivated by the studies in the Chern-Simons matrix models of the $\widehat D$ quiver \cite{CHJ,ADF,MN4}.} that the partition function of the orthosymplectic Chern-Simons matrix model is nothing but the chiral projection of the Chern-Simons matrix model with the super unitary gauge group
(see table 1 in \cite{MN5} for an interesting pattern).

\begin{figure}[!ht]
\centering\includegraphics[scale=1]{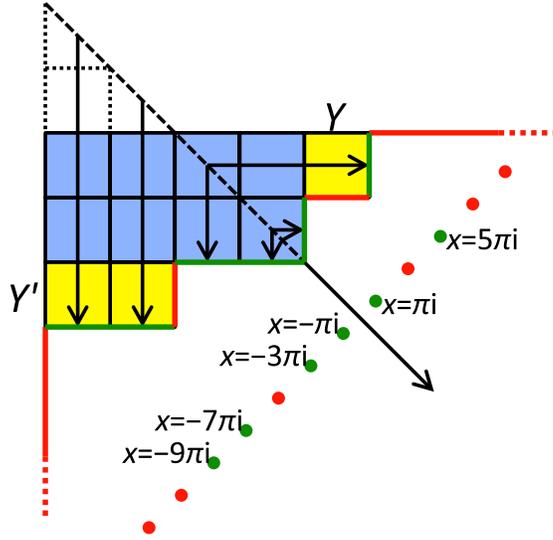}
\caption{Young diagram and Frobenius notation.
The standard Frobenius notation is defined by counting the boxes from the diagonal line.
For the super case U$(N_1|N_2)$, we shift the diagonal line by $M=N_2-N_1$.
For the Young diagram in the figure $[\alpha_1,\alpha_2,\alpha_3]=[5,4,2]$ or $[\lambda_1,\lambda_2,\lambda_3,\lambda_4,\lambda_5]^\text{T}=[3,3,2,2,1]^\text{T}$, the standard Frobenius notation is $(42|21)$ and the shifted one with $M=2$ is $(20|4310)$.
Here we find it useful to define the arm and leg lengths by adding $\frac{1}{2}$, or in other words, measuring the distances between the midpoints of two segments,
$(a_1,a_2|l_1,l_2,l_3,l_4)=(\frac{5}{2},\frac{1}{2}|\frac{9}{2},\frac{7}{2},\frac{3}{2},\frac{1}{2})$.
We sometimes decompose the Young diagram as $([(M+r)^r]+Y)\cup Y'$, which in the current case is $([4^2]+[1])\cup[2]$.
We also display the poles of $(2\cosh\frac{x}{2})^{-1}$ in \eqref{V} in our Fermi gas formalism \eqref{fermigas}.
The green dots denote the poles at $x=2\pi ia$ and $x=-2\pi il$ which are cancelled by the hyperbolic tangent functions in \eqref{V} and hence harmless, while the red dots denote the real poles which are not cancelled.}
\label{young}
\end{figure}

There is another formalism to study the matrix model \cite{sho} called ``open string formalism''.
Here we do not change the expression of the density matrix from the $M=0$ case and instead introduce many extra contributions with endpoints.
In this sense, as the Wilson loop expectation values \cite{HHMO}, we express the fractional brane background with the open string endpoints.
At present we stress that the open string formalism seems superior to the closed string one because it is obtained only from the combinatorics and hence is applicable not only for the partition function but also with the Wilson loop insertion.
In this formalism the expectation values of the half-BPS Wilson loop in the grand canonical ensemble\footnote{The integer $r$ (satisfying $0\le r\le N$) is defined later in \eqref{armlegdef} for a general Young diagram $Y$.
It is known that the supersymmetric Schur polynomial $s_Y(x|y)$ is vanishing for $N<r$.}
\begin{align}
\langle s_Y\rangle^\text{GC}_{k,M}(z)
=\sum_{N=r}^\infty z^{N-r}
\langle s_Y\rangle_{k,M}(N),
\label{sYvevGC}
\end{align}
is reduced to
\begin{align}
\frac{\langle s_Y\rangle^\text{GC}_{k,M}(z)}{\langle 1\rangle^\text{GC}_{k,0}(z)}
=\det\begin{pmatrix}\bigl(H_{l_p,-M+q-\frac{1}{2}}(z)\bigr)_{(M+r)\times M}&
\bigl(\widetilde H_{l_p,a_q}(z)\bigr)_{(M+r)\times r}\end{pmatrix},
\label{open}
\end{align}
where both $H_{l,a}(z)$ and $\widetilde H_{l,a}(z)$ take the form of a certain matrix element of $[1+z\widehat\rho_{M=0}]^{-1}$ (see \cite{sho} for the explicit form\footnote{We have slightly changed the notation from \cite{sho}.
In addition to changing the definition of the arm and leg lengths by $1/2$ as explained later in \eqref{armlegdef}, we also drop the overall factor $z$ from $\widetilde H_{l,a}(z)$.}).
The indices $a_q$, $l_p$ in \eqref{open} are the arm lengths and the leg lengths appearing in the Frobenius notation, which is another description of the Young diagram usually described by listing all of the arm lengths $[\alpha_1,\alpha_2,\cdots]$ or the leg lengths $[\lambda_1,\lambda_2,\cdots]^\text{T}$,
\begin{align}
&a_q=\alpha_q-q-M+1/2,\quad
l_p=\lambda_p-p+M+1/2,\nonumber\\
&r=\max\{q|a_q>0\}=\max\{p|l_p>0\}-M.
\label{armlegdef}
\end{align}
See figure \ref{young} for a pictorial explanation of the Frobenius notation.
Note that here we have deliberately added $\frac{1}{2}$ to the lengths to measure the distances between the midpoints of two segments.
As stressed in \cite{HHMO,HaOk}, the integrations in $H_{l,a}(z)$ and $\widetilde H_{l,a}(z)$ are convergent only for $a_1+l_1<k/2$ (which implies $M\le k/2$).
We shall follow this condition in our analysis.
One advantage of the open string formalism \eqref{open} is that we can prove the Giambelli compatibility for $\langle s_Y\rangle^\text{GC}_{k,M}(z)$ generally when correctly normalized with $\langle 1\rangle^\text{GC}_{k,M}(z)$ \cite{satsuki}.

Although many identities were proved in this context, still a lot of important duality relations await to be proved.
One of them is the miraculous open-closed duality observed recently in \cite{HaOk}.
In \cite{HaOk}, motivated by \cite{ADKMV}, starting from the simplest case with $r=1,M=0$, the authors arrive at a more general relation\footnote{In \cite{HaOk} the absolute values were taken for the expectation values in defining the grand canonical ensemble \eqref{sYvevGC}.
Hence, strictly speaking, the duality relation found in \cite{HaOk} is a consequence of \eqref{openclosed}.
Similarly, the relation \eqref{hook} also needs the modification of the complex conjugation.}
\begin{align}
\langle s_{[(M+r)^r]}\rangle^\text{GC}_{k,M}(z)
\sim\langle 1\rangle^\text{GC}_{k,M+2r}(z),
\label{openclosed}
\end{align}
with numerical computations.
Here $\sim$ means that the relation holds up to a numerical factor independent of $z$.
This duality relates the closed string BPS indices to the open string BPS indices and is another realization of the spirit of the open string formalism \cite{sho}, which expresses the closed string background formed by fractional branes with many open strings in the determinant.
In the same paper, the authors also observe a relation
\begin{align}
\langle s_{(a|l)}\rangle^\text{GC}_{k,M}(z)
\sim[\langle s_{(l+M|a-M)}\rangle^\text{GC}_{k,M}(z)]^*,
\label{hook}
\end{align}
for the hook representation $(a|l)$ with $a>M$.
The complex conjugation applies only for the coefficients of $z$.

In this paper, we shall generalize the closed string formalism \eqref{pf}, so that it incorporates the Wilson loop insertion.
Namely, contrary to the open string formalism \cite{sho} where we describe the fractional brane as a composite of the Wilson loops, here we propose an opposite formalism, which describes the Wilson loop by changing the closed string backgrounds,
\begin{align}
\frac{\langle s_Y\rangle_{k,M}(N)}{\langle s_Y\rangle_{k,M}(r)}
=\frac{1}{(N-r)!}\int\frac{d^{N-r}x}{(4\pi k)^{N-r}}
\prod_{i<j}^{N-r}\biggl(\tanh\frac{x_i-x_j}{2k}\biggr)^2
\prod_{i=1}^{N-r}V(x_i),
\label{fermigas}
\end{align}
with
\begin{align}
V(x)=\frac{1}{2\cosh\frac{x}{2}}
\prod_{a\in A}\tanh\frac{x-2\pi ia}{2k}
\prod_{l\in L}\tanh\frac{x+2\pi il}{2k}.
\label{V}
\end{align}
Here $A$ and $L$ denote respectively the set of all arm lengths and all leg lengths of the Young diagram $Y$.
Note that $\overline L$ appearing in \eqref{pfV} is the set of all leg lengths in the trivial representation.
From \eqref{V} it is easy to observe that although $x\in 2\pi i({\mathbb Z}\pm\frac{1}{2})$ are poles potentially, poles at $x=2\pi ia$ and $x=-2\pi il$ are cancelled by the zeros of the hyperbolic tangent functions (see figure \ref{young}).
Also, since $a_1+l_1<k/2$, a hyperbolic tangent does not induce a new pole at the zero of another hyperbolic tangent.

Using this new formalism we are able to prove some untouched dualities without difficulties.
Here we prove a generalized open-closed duality\footnote{The generalization without $Y$ was already observed in \cite{HaOk}.}
\begin{align}
\langle s_{([(M+r)^r]+Y)\cup Y'}\rangle^\text{GC}_{k,M}(z)
\sim\langle s_{([(M+r-1)^{r+1}]+Y)\cup Y'}\rangle^\text{GC}_{k,M-2}(z).
\label{generalized}
\end{align}
In fact, from the expression \eqref{fermigas} it is not difficult to find that we can shift the integration contour as long as we do not cross the poles.
Due to this reason, the expectation values of two Young diagrams in the grand canonical ensemble, which share the same set of
\begin{align}
\{a_q|q=1,\cdots,r\}\cup\{-l_p|p=1,\cdots,M+r\},
\label{A-L}
\end{align}
up to an overall shift of an integer, are identical.

This paper is organized as follows.
In the next section, we propose a new formalism to study the Wilson loop expectation values.
In section \ref{openclosedduality}, we use this formalism to prove (a generalized version of) the open-closed duality.
Finally in section \ref{discuss}, we conclude with discussions on the future directions.

\section{Wilson loop as closed string background}

In this section, following the method developed in \cite{MS2,MN5}, we shall rewrite the expectation value of the half-BPS Wilson loop \eqref{vev} into the partition function of a new Fermi gas system, where the density matrix is modified while the determinant structure is kept fixed.
We shall rescale the integration variables by $k$ as
\begin{align}
\langle s_Y\rangle_{k,M}(N)
&=\frac{(-1)^{\frac{1}{2}N_1(N_1-1)+\frac{1}{2}N_2(N_2-1)}}{N_1!N_2!}
\int\frac{d^{N_1}\mu}{\hbar^{N_1}}\frac{d^{N_2}\nu}{\hbar^{N_2}}
e^{\frac{i}{2\hbar}(\sum_{m=1}^{N_1}\mu_m^2-\sum_{n=1}^{N_2}\nu_n^2)}\nonumber\\
&\qquad\times\biggl[\frac{\prod_{m<m'}^{N_1}2\sinh\frac{\mu_m-\mu_{m'}}{2k}
\prod_{n<n'}^{N_2}2\sinh\frac{\nu_n-\nu_{n'}}{2k}}
{\prod_{m=1}^{N_1}\prod_{n=1}^{N_2}2\cosh\frac{\mu_m-\nu_n}{2k}}\biggr]^2
s_Y(e^{\frac{\mu}{k}}|e^{\frac{\nu}{k}}),
\end{align}
with $\hbar=2\pi k$ and $k>0$.
We begin our analysis by assuming $M=N_2-N_1\ge 0$ and denote $N_1=N$, $N_2=M+N$ and $\langle s_Y\rangle_k(N,M+N)=\langle s_Y\rangle_{k,M}(N)$.

As in the open string formalism \cite{sho}, our starting point is the following three determinant formulas; the Cauchy-Vandermonde determinant
\begin{align}
\frac{\prod_{m<m'}^{N_1}(x_m-x_{m'})
\prod_{n<n'}^{N_2}(y_n-y_{n'})}
{\prod_{m=1}^{N_1}\prod_{n=1}^{N_2}(x_m+y_n)}
=(-1)^{N_1(N_2-N_1)}
\det\begin{pmatrix}\biggl[\displaystyle\frac{1}{x_m+y_n}\biggr]
_{(m,n)\in Z_{1}\times Z_{2}}\\
\Bigl[y_n^{\overline l-\frac{1}{2}}\Bigr]
_{(\overline l,n)\in\overline L\times Z_{2}}
\end{pmatrix},
\end{align}
where $m$, $n$ and $\overline l$ are respectively elements of $Z_1=\{1,2,\cdots,N_1\}$, $Z_2=\{1,2,\cdots,N_2\}$ and $\overline L=\{M-\frac{1}{2},M-\frac{3}{2},\cdots,\frac{1}{2}\}$ in this order;
the same determinant
\begin{align}
&(-1)^{\frac{1}{2}N_1(N_1-1)+\frac{1}{2}N_2(N_2-1)}
\frac{\prod_{m<m'}^{N_1}(x_{m'}^{-1}-x_m^{-1})
\prod_{n<n'}^{N_2}(y_{n'}^{-1}-y_n^{-1})}
{\prod_{m=1}^{N_1}\prod_{n=1}^{N_2}(x_m^{-1}+y_n^{-1})}\nonumber\\
&=(-1)^{N_1(N_2-N_1)}
\det\begin{pmatrix}\biggl[\displaystyle\frac{1}{y_n^{-1}+x_m^{-1}}\biggr]
_{(n,m)\in Z_{2}\times Z_{1}}&
\Bigl[y_n^{-\overline l+\frac{1}{2}}\Bigr]
_{(n,\overline l)\in Z_{2}\times\overline L}
\end{pmatrix},
\end{align}
obtained by the substitutions $x_m\to x_m^{-1}$ and $y_n\to y_n^{-1}$;
the determinantal formula for the supersymmetric Schur polynomial due to Moens and Van der Jeugt \cite{MVdJ}
\begin{align}
\frac{s_Y(x|y)}{(-1)^r}
=\det\begin{pmatrix}\biggl[\displaystyle\frac{1}{x_m+y_n}\biggr]
_{(m,n)\in Z_{1}\times Z_{2}}&
\Bigl[x_m^{a-\frac{1}{2}}\Bigr]
_{(m,a)\in Z_{1}\times A}\\
\Bigl[y_n^{l-\frac{1}{2}}\Bigr]
_{(l,n)\in L\times Z_{2}}&
[0]_{L\times A}\end{pmatrix}\bigg/
\det\begin{pmatrix}\biggl[\displaystyle\frac{1}{x_m+y_n}\biggr]
_{(m,n)\in Z_{1}\times Z_{2}}\\
\Bigl[y_n^{\overline l-\frac{1}{2}}\Bigr]
_{(\overline l,n)\in\overline L\times Z_{2}}\end{pmatrix},
\end{align}
with $A=\{a_1,a_2,\cdots,a_r\}$ and $L=\{l_1,l_2,\cdots,l_{M+r}\}$, where, as in figure \ref{young}, the arm and leg lengths measure the distances between the midpoints of two segments and are greater than the standard ones by $\frac{1}{2}$.
After multiplying these three formulas with the substitutions $x_m=e^{\frac{\mu_m}{k}}$ and $y_n=e^{\frac{\nu_n}{k}}$, we find
\begin{align}
&(-1)^{\frac{1}{2}N_1(N_1-1)+\frac{1}{2}N_2(N_2-1)}
\biggl[\frac{\prod_{m<m'}^{N_1}2\sinh\frac{\mu_m-\mu_{m'}}{2k}
\prod_{n<n'}^{N_2}2\sinh\frac{\nu_n-\nu_{n'}}{2k}}
{\prod_{m=1}^{N_1}\prod_{n=1}^{N_2}2\cosh\frac{\mu_m-\nu_n}{2k}}\biggr]^2
\frac{s_Y(e^{\frac{\mu}{k}}|e^{\frac{\nu}{k}})}{(-1)^r}\nonumber\\
&=\det\begin{pmatrix}
\biggl[\displaystyle\frac{1}{2\cosh\frac{\mu_m-\nu_n}{2k}}\biggr]_{N_1\times N_2}&
\bigl[e^{\frac{a\mu_m}{k}}\bigr]_{N_1\times r}\\
\bigl[e^{\frac{l\nu_n}{k}}\bigr]_{(M+r)\times N_2}&
[0]_{(M+r)\times r}
\end{pmatrix}
\det\begin{pmatrix}
\biggl[\displaystyle\frac{1}{2\cosh\frac{\nu_n-\mu_m}{2k}}\biggr]_{N_2\times N_1}
&\bigl[e^{-\frac{\overline l\nu_n}{k}}\bigr]_{N_2\times M}
\end{pmatrix}.
\end{align}

Next, let us rewrite the components of the determinants by using the Fourier transformation of $(2\cosh\frac{\widehat p}{2})^{-1}$ and introducing the formal states $|h\rrangle$ and $\llangle h|$ as in \cite{MS2}, $(h\in{\mathbb Z}\pm\frac{1}{2})$
\begin{align}
\langle\mu|\frac{1}{2\cosh\frac{\widehat p}{2}}|\nu\rangle
=\frac{1}{k}\frac{1}{2\cosh\frac{\mu-\nu}{2k}},\qquad
\llangle h|\mu\rangle=\langle\mu|h\rrangle=e^{\frac{h\mu}{k}}.
\end{align}
Then, we can follow the standard tricks of including the Gaussian factor into the brackets and applying the similarity transformation
\begin{align}
1=\int\frac{dq}{2\pi}|q\rangle\langle q|\quad\Rightarrow\quad
1=\int\frac{dq}{2\pi}e^{-\frac{i}{2\hbar}\widehat p^2}|q\rangle
\langle q|e^{\frac{i}{2\hbar}\widehat p^2},
\end{align}
to the $\mu$ and $\nu$ integrations.
After these manipulations, the expectation value is expressed as
\begin{align}
&\langle s_Y\rangle_{k,M}(N)
=\frac{(-1)^r}{N_1!N_2!}\int\frac{d^{N_1}\mu}{\hbar^{N_1}}
\frac{d^{N_2}\nu}{\hbar^{N_2}}\nonumber\\
&\quad\times\det\begin{pmatrix}
\biggl[\displaystyle k\langle\mu_m|e^{\frac{i}{2\hbar}\widehat p^2}
e^{\frac{i}{2\hbar}\widehat q^2}\frac{1}{2\cosh\frac{\widehat p}{2}}
e^{-\frac{i}{2\hbar}\widehat q^2}e^{-\frac{i}{2\hbar}\widehat p^2}|\nu_n\rangle
\biggr]_{N_1\times N_2}
&\Bigl[\langle\mu_m|e^{\frac{i}{2\hbar}\widehat p^2}
e^{\frac{i}{2\hbar}\widehat q^2}|a\rrangle\Bigr]_{N_1\times r}\\
\Bigl[\llangle l|e^{-\frac{i}{2\hbar}\widehat q^2}
e^{-\frac{i}{2\hbar}\widehat p^2}|\nu_n\rangle\Bigr]_{(M+r)\times N_2}&
\bigl[0\bigr]_{(M+r)\times r}
\end{pmatrix}\nonumber\\
&\quad\times\det\begin{pmatrix}
\biggl[\displaystyle k\langle\nu_n|e^{\frac{i}{2\hbar}\widehat p^2}
\frac{1}{2\cosh\frac{\widehat p}{2}}e^{-\frac{i}{2\hbar}\widehat p^2}
|\mu_m\rangle\biggr]_{N_2\times N_1}&
\Bigl[\langle\nu_n|e^{\frac{i}{2\hbar}\widehat p^2}
|{-\overline l}\rrangle\Bigr]_{N_2\times M}
\end{pmatrix}.
\end{align}
It is magical \cite{MS2} that all of the components in the first determinant reduce to delta functions\footnote{The formal computation in the second and third formulas needs justification \cite{MS2}.
It is important that the remaining factors $(2\cosh\frac{\mu-\nu}{2k})^{-1}$ do not contain poles between $\im\mu=2\pi a_1$ and $\im\nu=-2\pi l_1$ due to the condition $a_1+l_1<k/2$.
We are grateful to Takao Suyama for valuable discussions.}
\begin{align}
k\langle\mu|e^{\frac{i}{2\hbar}\widehat p^2}e^{\frac{i}{2\hbar}\widehat q^2}
\frac{1}{2\cosh\frac{\widehat p}{2}}
e^{-\frac{i}{2\hbar}\widehat q^2}e^{-\frac{i}{2\hbar}\widehat p^2}|\nu\rangle
&=\frac{\hbar}{2\cosh\frac{\mu}{2}}\delta(\mu-\nu),
\nonumber\\
\langle\mu|e^{\frac{i}{2\hbar}\widehat p^2}
e^{\frac{i}{2\hbar}\widehat q^2}|a\rrangle
&=\frac{\hbar}{\sqrt{-ik}}e^{\frac{i}{2\hbar}(2\pi a)^2}\delta(\mu-2\pi ia),
\nonumber\\
\llangle l|e^{-\frac{i}{2\hbar}\widehat q^2}
e^{-\frac{i}{2\hbar}\widehat p^2}|\nu\rangle
&=\frac{\hbar}{\sqrt{ik}}e^{-\frac{i}{2\hbar}(2\pi l)^2}\delta(\nu+2\pi il).
\end{align}
For the second determinant we have
\begin{align}
\langle\nu|e^{\frac{i}{2\hbar}\widehat p^2}|{-\overline l}\rrangle
&=e^{-\frac{i}{2\hbar}(2\pi\overline l)^2}
\langle\nu|{-\overline l}\rrangle.
\end{align}
For the expansion of the first determinant to be non-vanishing we need to choose $r$ rows out of $N$ rows in the upper-right block and $M+r$ columns out of $M+N$ columns in the lower-left block.
Then, the remaining $N-r$ components are chosen from the upper-left block.
In renaming the indices, we have $r!(M+r)!(N-r)!$ identical terms with signs $(-1)^{(M+r)r}$.
After combining these factors, we find
\begin{align}
&\langle s_Y\rangle_{k,M}(N)=\frac{(-1)^{\frac{1}{2}M(M-1)+Mr}}{(N-r)!}
\int\frac{d^N\mu}{\hbar^N}\frac{d^{M+N}\nu}{\hbar^{M+N}}
e^{\frac{i}{2\hbar}(2\pi)^2(\sum a^2-\sum l^2-\sum\overline l^2)}\nonumber\\
&\qquad\times
\prod_{i=1}^{N-r}\frac{\hbar}{2\cosh\frac{\mu_i}{2}}\delta(\mu_i-\nu_i)
\prod_{q=1}^r\frac{\hbar}{\sqrt{-ik}}\delta(\mu_{N-r+q}-2\pi ia_q)
\prod_{p=1}^{M+r}\frac{\hbar}{\sqrt{ik}}\delta(\nu_{N-r+p}+2\pi il_p)
\nonumber\\
&\qquad\times
e^{\frac{M}{2k}(\sum_{m=1}^N\mu_m-\sum_{n=1}^{M+N}\nu_n)}
\frac{\prod_{m<m'}^{N}2\sinh\frac{\mu_m-\mu_{m'}}{2k}
\prod_{n<n'}^{M+N}2\sinh\frac{\nu_n-\nu_{n'}}{2k}}
{\prod_{m=1}^N\prod_{n=1}^{M+N}2\cosh\frac{\mu_m-\nu_n}{2k}},
\end{align}
where we have used $(-1)^{r^2}=(-1)^r$.
After performing the integration of the delta functions by substitutions, we arrive at the expression
\begin{align}
&\frac{\langle s_Y\rangle_{k,M}(N)}
{\langle s_Y\rangle_{k,M}(r)}
=\frac{1}{(N-r)!}\int\frac{d^{N-r}x}{(4\pi k)^{N-r}}
\prod_{i<j}^{N-r}\biggl(\tanh\frac{x_i-x_j}{2k}\biggr)^2\nonumber\\
&\qquad\times\prod_{i=1}^{N-r}\biggl[\frac{1}{2\cosh\frac{x_i}{2}}
\prod_{a\in A}\tanh\frac{x_i-2\pi ia}{2k}
\prod_{l\in L}\tanh\frac{x_i+2\pi il}{2k}\biggr],
\label{closed}
\end{align}
where the normalization is given by\footnote{Note that $N=r$ is the smallest case for the expectation value $\langle s_Y\rangle_{k,M}(N)$ to be non-vanishing.
The absolute value $|\langle s_Y\rangle_{k,M}(r)|$ is coincident with $C_Y(k,M)$ in \cite{HaOk}.}
\begin{align}
\langle s_Y\rangle_{k,M}(r)
&=\frac{i^{\frac{1}{2}M(M-1)+Mr}
e^{\frac{\pi i}{k}(\sum a^2-\sum l^2-\sum\overline l^2)}
e^{\frac{\pi i}{k}M(\sum a+\sum l)}}{\sqrt{-ik}^r\sqrt{ik}^{M+r}}
\frac{\prod_{a>a'}2\sin\frac{\pi(a-a')}{k}
\prod_{l>l'}2\sin\frac{\pi(l-l')}{k}}
{\prod_a\prod_l2\cos\frac{\pi(a+l)}{k}}.
\label{rclosed}
\end{align}

We note that, although we originally start with the situation of $N_1\le N_2$, both of the formulas \eqref{closed} and \eqref{rclosed} are valid for the opposite case $N_1>N_2$ as well\footnote{
Instead of the leg lengths for the trivial representation $\overline L$, we introduce the arm lengths $\overline a\in\overline A=\{-M-\frac{1}{2},-M-\frac{3}{2},\cdots,\frac{1}{2}\}$ for the case $M<0$.
Hence, we need to interpret the phase factor $e^{\frac{\pi i}{k}(-\sum\overline l^2)}$ in \eqref{rclosed} as $e^{\frac{\pi i}{k}(+\sum\overline a^2)}$.
} if we stick to the original notation $N_1=N$, $N_2=M+N$, $\max\{q|a_q>0\}=r$, $\max\{p|l_p>0\}=M+r$ and $\langle s_Y\rangle_{k,M}(N)=\langle s_Y\rangle_k(N,M+N)$ but $M<0$.
Therefore, no special care is needed when crossing the diagonal line and we can extend to $M<0$ directly.

Finally let us comment on the closed string formalism for the partition function \eqref{pf}.
In \cite{Ho1} the proof was done by several steps of integrations.
Following the method of \cite{MS2,MN5}, we have generalized the formalism for the expectation values of the half-BPS Wilson loop.
The result of the partition function can be simply rederived by setting $A=\emptyset, L=L'$ in \eqref{closed}.

\section{Proof of generalized open-closed duality}\label{openclosedduality}

In the previous section we have found that, after suitably normalized, the expectation value $\langle s_Y\rangle_{k,M}(N)$ is given in \eqref{fermigas} with $V(x)$ defined by \eqref{V} and no special care is needed when crossing $M=0$.
From figure \ref{young} we know that $V(x)$ contains poles periodically, though some of them are cancelled by the zeros of the hyperbolic tangent functions.
Therefore, as long as we do not encounter the real poles, we can shift the integration contour or in contrast the position of the poles by $\pm 2\pi i$ freely, so that two expectation values which share the same set of \eqref{A-L} up to an integral shift are identical.
This identity induces the duality relation we want to prove.
Let us see this explicitly for the example in figure \ref{young}.
Here we assume that $k$ is large enough so that we do not have to consider the poles of the hyperbolic tangent functions in our shift of the integration contour.

\begin{figure}[ht!]
\centering\includegraphics[scale=1]{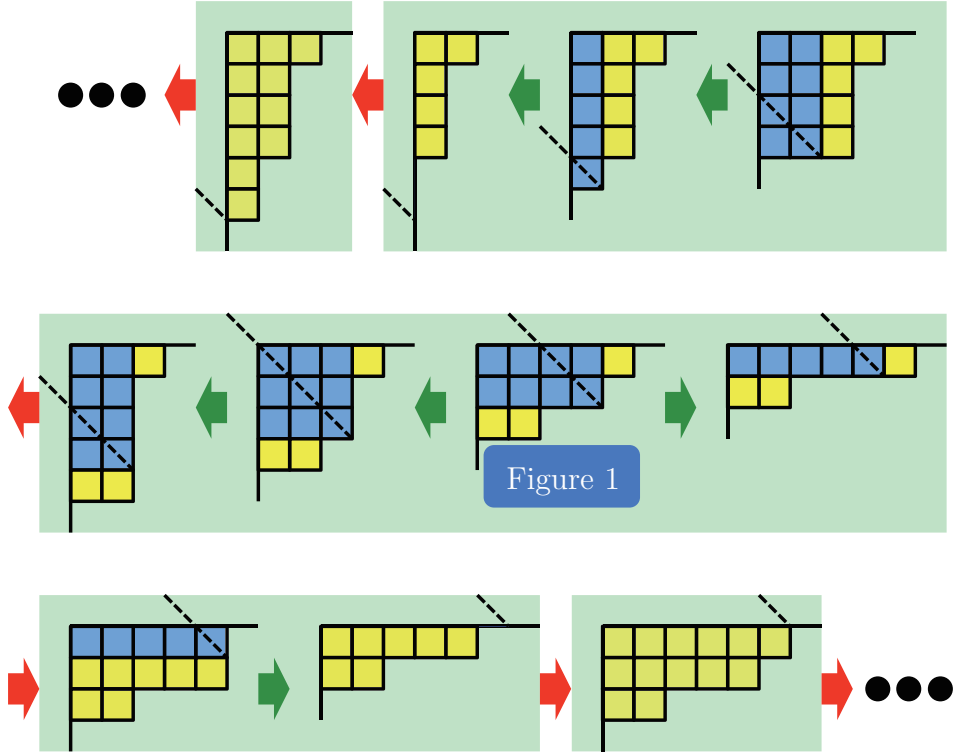}\\
\vspace{-41mm}\hspace{23mm}{\color{white}Figure \ref{young}}
\vspace{41mm}
\caption{Shifting the integration contour across the harmless poles does not change the expectation values and hence induces the duality, while shifting across the poles is not allowed.
We assume that $k$ is large enough.}
\label{shift}
\end{figure}

In figure \ref{shift} we pick up the example in figure \ref{young} and shift the integration contour in the unit of $2\pi i$.
Starting from $(\frac{5}{2},\frac{1}{2}|\frac{9}{2},\frac{7}{2},\frac{3}{2},\frac{1}{2})$, if we move the integration coutour upwards (so that the number of leg lengths increases), we find $(\frac{3}{2}|\frac{11}{2},\frac{9}{2},\frac{5}{2},\frac{3}{2},\frac{1}{2})$, where in this shift we only move across the harmless green pole ($x=\pi i$ in figure \ref{young}) and the expectation value is not changed.
If we move further into $(\frac{1}{2}|\frac{13}{2},\frac{11}{2},\frac{7}{2},\frac{5}{2},\frac{3}{2})$, now we need to cross the real red pole ($x=3\pi i$ in figure \ref{young}) and the expectation value is changed.
Similarly, we can move downwards to $(\frac{7}{2},\frac{3}{2},\frac{1}{2}|\frac{7}{2},\frac{5}{2},\frac{1}{2})$ and so on without changing the expectation values for a while.
We have classified the expectation values by the shaded green backgrounds in figure \ref{shift}.
We can summarize the above duality relation as
\begin{align}
\frac{\langle s_{([(M+r)^r]+Y)\cup Y'}\rangle^\text{GC}_{k,M}(z)}
{\langle s_{([(M+r)^r]+Y)\cup Y'}\rangle_{k,M}(r)}
=\frac{\langle s_{([(M+r-1)^{r+1}]+Y)\cup Y'}\rangle^\text{GC}_{k,M-2}(z)}
{\langle s_{([(M+r-1)^{r+1}]+Y)\cup Y'}\rangle_{k,M-2}(r+1)},
\end{align}
if $\lambda_1\le r$ and $\alpha_1\le M+r-1$ where $\lambda_1=\lambda_1(Y)$ and $\alpha_1=\alpha_1(Y')$ denote the first leg length of $Y$ and the first arm length of $Y'$.
Using this recursively, we find
\begin{align}
\frac{\langle s_{([(\widetilde M-\lambda_1)^{\lambda_1}]+Y)\cup Y'}\rangle
^\text{GC}_{k,\widetilde M-2\lambda_1}(z)}
{\langle s_{([(\widetilde M-\lambda_1)^{\lambda_1}]+Y)\cup Y'}\rangle
_{k,\widetilde M-2\lambda_1}(\lambda_1)}
=\frac{\langle s_{([(\alpha_1)^{\widetilde M-\alpha_1}]+Y)\cup Y'}\rangle
^\text{GC}_{k,-\widetilde M+2\alpha_1}(z)}
{\langle s_{([(\alpha_1)^{\widetilde M-\alpha_1}]+Y)\cup Y'}\rangle
_{k,-\widetilde M+2\alpha_1}(\widetilde M-\alpha_1)},
\label{max}
\end{align}
with $\widetilde M=M+2r$.
For the special case $\lambda_1=\alpha_1=0$ this reduces to \eqref{openclosed}.

In \cite{HaOk} another interesting identity \eqref{hook}
\begin{align}
\frac{\langle s_{(a|l)}\rangle^\text{GC}_{k,M}(z)}
{\langle s_{(a|l)}\rangle_{k,M}(1)}
=\biggl[\frac{\langle s_{(l+M|a-M)}\rangle^\text{GC}_{k,M}(z)}
{\langle s_{(l+M|a-M)}\rangle_{k,M}(1)}\biggr]^*,
\label{armleg}
\end{align}
with $a>M$ was found numerically.
However, after generalizing the open-closed duality into \eqref{max}, we point out that this falls into the same class of the identity.
Namely, since the Young diagram $(a|l)$ is decomposed as $([(M+1)^1]+[a-M-\frac{1}{2}])\cup[1^{l-\frac{1}{2}}]$, we can use our formula \eqref{max} to shift the integration contour to obtain $([1^{M+1}]+[a-M-\frac{1}{2}])\cup[1^{l-\frac{1}{2}}]$, which is $(a-M|l+M)$.
After applying the conjugate relation \eqref{conj}, this reduces to \eqref{armleg}.

Let us comment on the effect of crossing the real poles in shifting the integration contour.
Although this gives different values due to the effect of the poles, we note that the difference is under control by taking care of the residues.
The computation of the difference is similar to \cite{HHMO,sho,MS1}.

\section{Discussions}\label{discuss}

In this paper we have proposed a new Fermi gas formalism to study vacuum expectation values of the half-BPS Wilson loop.
Compared with the open string formalism \cite{sho}, which expresses the partition function with fractional branes using the Wilson loops, our formalism is the opposite of it.
Namely, with the same method which leads to the closed string formalism of the partition function, we end up with a new formalism that expresses the Wilson loop expectation values using the modified density matrix which depends on the set of arm lengths and leg lengths.
With this formalism we can prove some important duality relations proposed previously: the open-closed duality and its generalizations.

We comment that our formalism looks similar to the one conjectured in \cite{HNS}, though the comparison seems difficult.
Also, although we have proved the identity \eqref{generalized} inspiring the open-closed duality, it is not clear to us how this duality relates to those in \cite{GV,OV,ADKMV}.
It would be interesting to clarify the relations\footnote{We are grateful to Shinji Hirano, Takahiro Nishinaka and Masaki Shigemori for valuable discussions.}.

The computation of the half-BPS trace operators in D3-branes \cite{CJR} has a nice interpretation from the fermion droplets \cite{LLM}.
After obtaining a simple formalism for the half-BPS Wilson loop in M2-branes, it is interesting to ask whether we can find a similar interpretation from the supergravity viewpoint as well.

Unfortunately, because of the convergence condition $M\le k/2$, our formalism seems not very helpful in proving the Giveon-Kutasov duality \cite{GK,KWYdual} relating $M$ to $k-M$.
In fact, previously the duality was used to define the formalism for $M\ge k/2$.
It is an interesting open problem to improve this situation.

From a technical viewpoint, we have proposed another nice formalism for general expectation values in the ABJM matrix model.
Since there are some interesting related models \cite{GHN,MN1,MN2,MN3,HHO}, we would like to apply a similar formalism to these models for the numerical computations and find more relations to the topological string theories.

\section*{Acknowledgements}
We are grateful to Heng-Yu Chen, Shun-Jen Cheng, Shinji Hirano, Pei-Ming Ho, Yoshinori Honma, Dharmesh Jain, Takuya Matsumoto, Satsuki Matsuno, Shota Nakayama, Takahiro Nishinaka, Tadashi Okazaki, Takeshi Oota, Masaki Shigemori, Reiji Yoshioka and especially Takao Suyama for valuable discussions.
The work of S.M.\ is supported by JSPS Grant-in-Aid for Scientific
Research (C) \# 26400245.

\end{document}